\documentclass[prl,amsmath,amssymb,showpacs,superscriptaddress,twocolumn]
{revtex4}
\usepackage{longtable,dcolumn,epsfig,color}

\begin{document} 

\title{Low-energy enhancement of magnetic dipole radiation}

\author{R. Schwengner} 
\affiliation{Institute of Radiation Physics, Helmholtz-Zentrum
             Dresden-Rossendorf, 01328 Dresden, Germany}
\author{S. Frauendorf}
\affiliation{Department of Physics, University of Notre Dame,
             Notre Dame, Indiana 46556, USA}
\author{A. C. Larsen}
\affiliation{Department of Physics, University of Oslo, 0316 Oslo, Norway}

\date{\today} 

\begin{abstract}
Magnetic dipole strength functions have been deduced from averages of a large
number of $M1$ transition strengths calculated within the shell model for the
nuclides $^{90}$Zr, $^{94}$Mo, $^{95}$Mo, and $^{96}$Mo. An enhancement of $M1$
strength toward low transition energy has been found for all nuclides
considered. Large $M1$ strengths appear for transitions between close-lying
states with configurations including proton as well as neutron high-$j$ orbits
that re-couple their spins and add up their magnetic moments coherently. The
$M1$ strength function deduced from the calculated $M1$ transition strengths
is compatible with the low-energy enhancement found in ($^3$He,$^3$He') and
$(d,p)$ experiments. The present work presents for the first time an
explanation of the experimental findings.
\end{abstract}


\maketitle

Photonuclear reactions and the inverse radiative-capture reactions between
nuclear states in the region of high excitation energy and large level density,
the so-called quasicontinuum of states, are of considerable interest in many
applications. Radiative neutron capture, for example, plays a central role in
the synthesis of the elements in various stellar environments
\cite{Arnould07,Kaeppler11}. An improved theoretical description of neutron
capture reactions is important for next-generation nuclear technologies, such
as the transmutation of long-lived nuclear waste \cite{Arnould07,Chadwick11}.
Rates of these reactions are calculated using codes that are based on the
statistical reaction theory (e.g. TALYS \cite{kon05}). A critical input to
these calculations is the average electromagnetic transition strengths,
described by photon strength functions. For example, modifications of the
electric-dipole ($E1$) strength function can cause drastic changes in the
abundances of elements produced via neutron capture in the r-process occurring
in violent stellar events \cite{gor98}.

In the energy range below about 10 MeV, which is relevant for the applications,
the dipole strength function $f_1$ is dominated by the tail of the isovector
electric giant dipole resonance (GDR), which is the collective vibration of the
neutron system against the proton system. The damped vibration is described by
a Lorentz shape to $f_1(E_\gamma)$ \cite{bri55,axe62,cap10}, where $E_\gamma$
is the energy of the photon. Combinations of two or three Lorentz curves are
used to describe the double or triple humps of the GDR caused by quadrupole and
triaxial deformation of the nuclei \cite{boh75,eis75,jun08}. Such a
parametrization gives a good description of the experimental photoabsorption
cross section $\sigma_\gamma$ = 3 $(\pi \hbar c)^2$ $E_\gamma$ $f_1(E_\gamma)$
of nuclei in the ground state. It is quite common to adopt the so-called
Brink-Axel hypothesis \cite{bri55,axe62}, which states that the strength
function does not depend on the excitation energy. This means that the same
strength function describes the emission of photons from highly exctited
states, following e.g. neutron capture. The Generalized Lorentzian (GLO)
\cite{kop90} includes a correction to the Standard Lorentzian (SLO)
\cite{bri55,axe62}, which accounts for the temperature of the nucleus
emitting the photons.

For the magnetic dipole $(M1)$ contribution to $f_1$, two types of excitations
have been considered so far. The scissors mode, which is interpreted as a
small-amplitude rotation of the neutron system against the proton system,
generates a bump of the $M1$ strength around 3 MeV in deformed nuclei
\cite{hey10}. After it had been well established  in the absorption spectra of
the ground state, it was recently also identified in the emission from highly
excited states (see Ref.~\cite{gut12} and earlier work cited therein).
At higher energy, typically around 8 MeV, the $M1$ strength is dominated by the
spin-flip resonance \cite{hey10}. Phenomenological $M1$ strength functions used
in statistical-reaction codes are approximated by Lorentz curves with
parameters usually derived from systematics \cite{cap10}. 

\begin{figure}
\vspace*{-6cm}
\epsfig{file=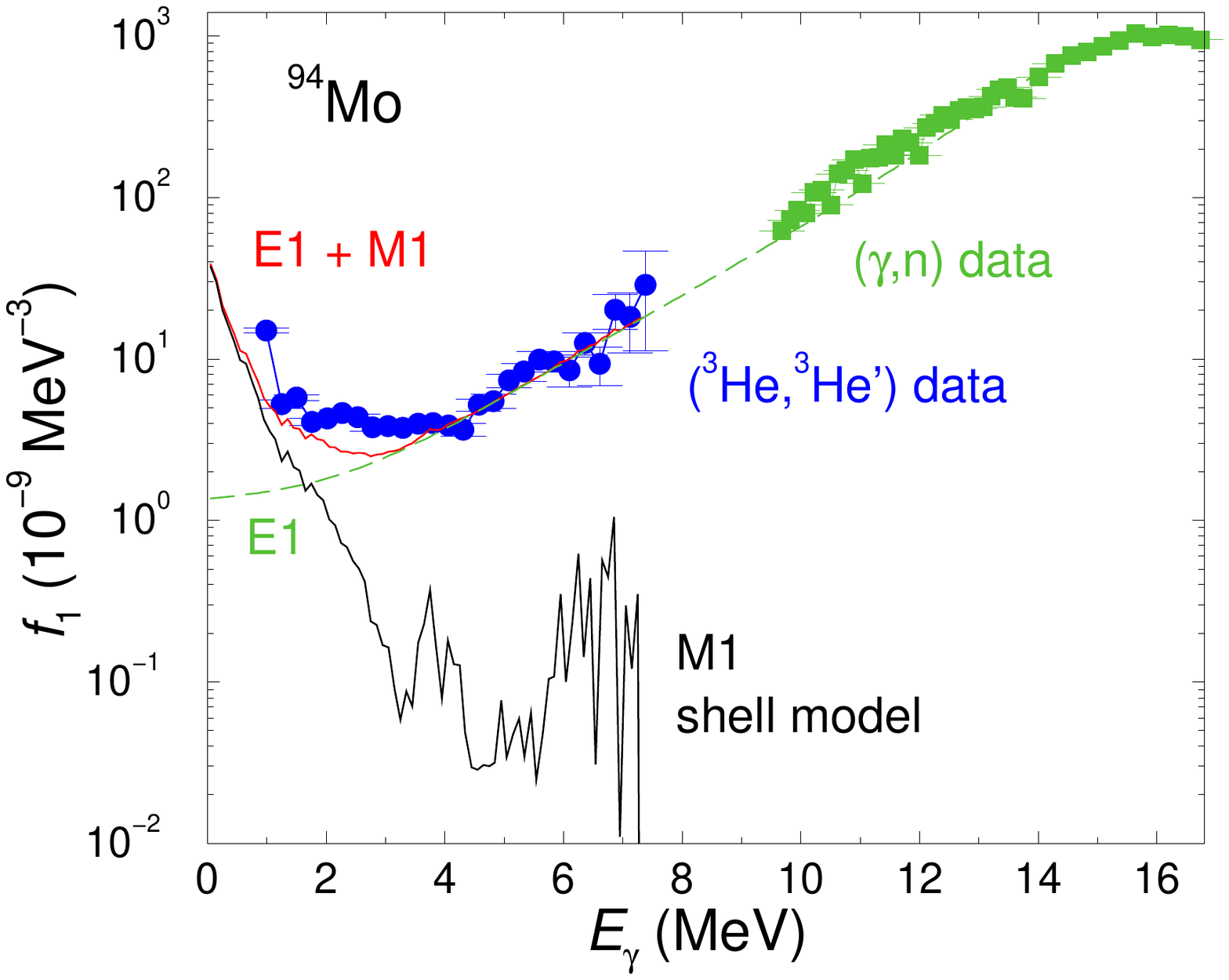,width=10cm}
\caption{\label{fig:94MoEgf1}(Color online) Strength functions for $^{94}$Mo 
deduced from ($^3$He,$^3$He') (blue circles) and $(\gamma,n)$ (green squares)
experiments, the $M1$ strength function from the present shell model
calculations (black solid line), $E1$ strength according to the GLO expression
with parameters $E_0$ = 16.36 MeV, $\sigma_0$ = 185 b, $\Gamma$ = 5.5 MeV,
$T$ = 0.35 MeV (green dashed line), and the total ($E1 + M1$) dipole
strength function (red line).}
\end{figure}

\begin{figure}
\vspace*{-6cm}
\epsfig{file=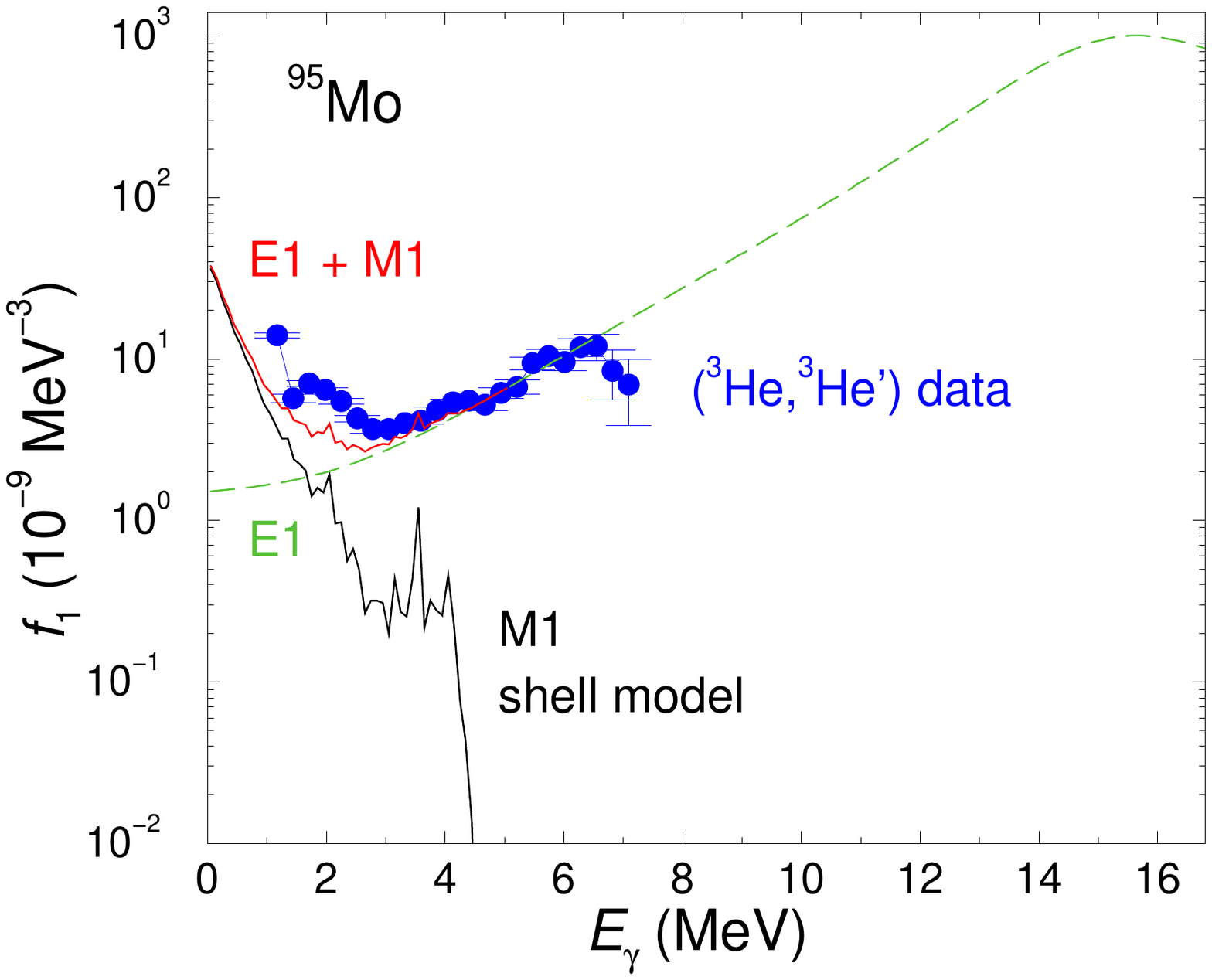,width=10cm}
\caption{\label{fig:95MoEgf1}(Color online) As Fig.~\ref{fig:94MoEgf1}, but for
$^{95}$Mo. $(\gamma,n)$ data are not available for $^{95}$Mo.} 
\end{figure}

\begin{figure}
\vspace*{-6cm}
\epsfig{file=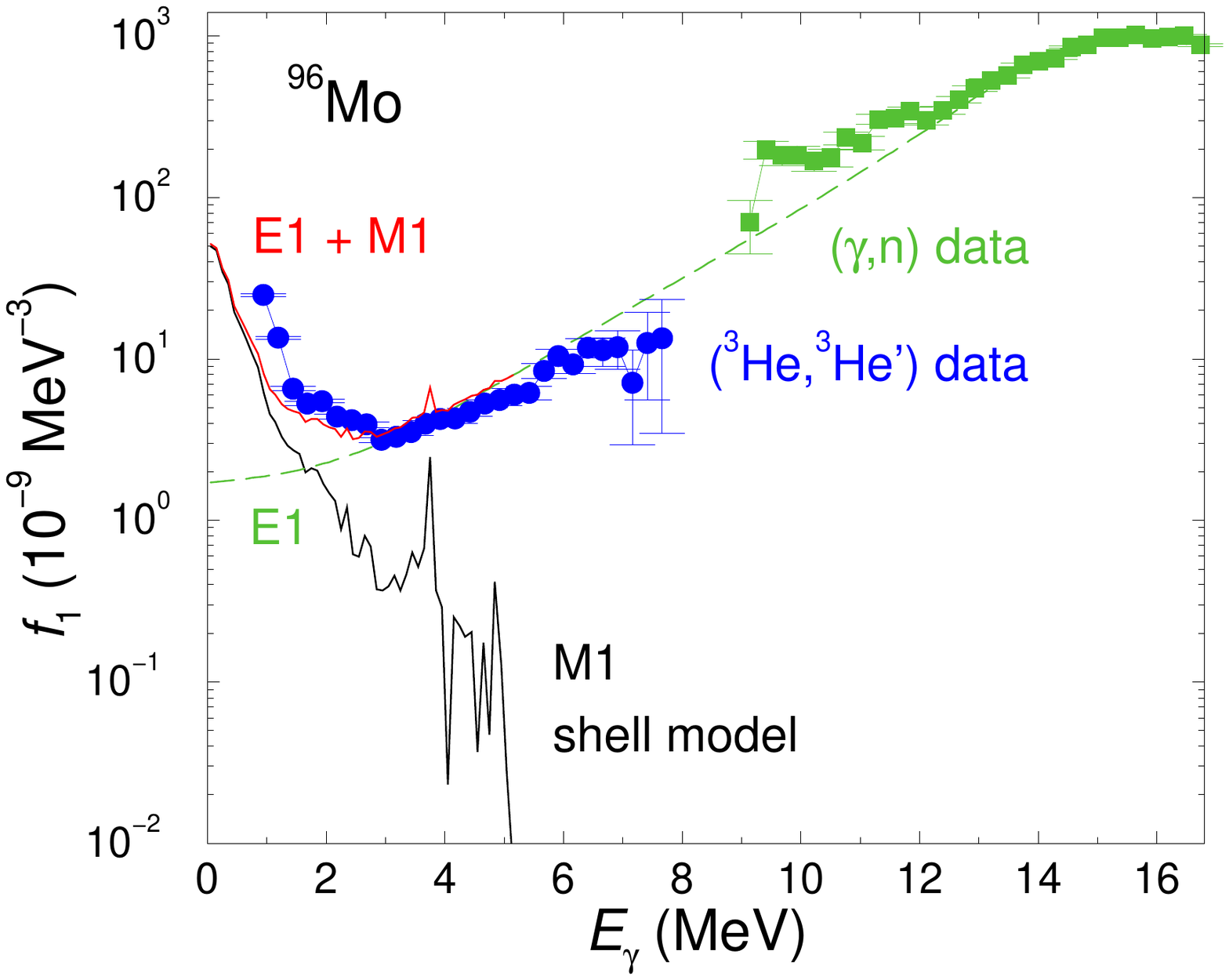,width=10cm}
\caption{\label{fig:96MoEgf1}(Color online) As Fig.~\ref{fig:94MoEgf1}, but for
$^{96}$Mo.}
\end{figure}

The Lorentz curves used for the $E1$ and $M1$ strength functions decrease when
approaching $E_\gamma$ = 0. In contrast, an increase of the dipole strength
function below 3 MeV toward low $\gamma$-ray energy has been found in several
nuclides in the mass range from $A \approx$ 50 to 100, such as $^{56,57}$Fe
\cite{voi04}, $^{60}$Ni \cite{voi10}, and $^{105,106}$Cd \cite{lar13}. In
particular, this low-energy enhancement of the strength function was deduced
from experiments using ($^3$He,$^3$He') reactions on various Mo isotopes
\cite{gut05} and was confirmed in an independent experiment using the
$^{94}$Mo$(d,p)^{95}$Mo reaction \cite{wie12}. The ($^3$He,$^3$He') data for
$^{94}$Mo, $^{95}$Mo, and $^{96}$Mo are shown in Figs.~\ref{fig:94MoEgf1},
\ref{fig:95MoEgf1}, and \ref{fig:96MoEgf1}, respectively, together with
$(\gamma,n)$ data \cite{bei74}. The increase at low $\gamma$-ray energies may
have a potentially large impact on neutron-capture reaction rates relevant for
astrophysical processes \cite{lar10}. Neither of these measurements were able
to distinguish between $E1$ and $M1$ strength. An indication for an $M1$
character of the low-energy enhancement was discussed for the case of $^{60}$Ni
\cite{voi10}. The $\gamma$-ray absorption from the ground state of even-even
nuclei leads to only few discrete 1$^+$ and 1$^-$ levels below 2 MeV.
Obviously, it does not make sense to invoke the Brink-Axel hypothesis to relate
these absorption cross sections to the observed low-energy enhancement of the
$\gamma$-ray emission from highly excited levels. The properties of the $E1$
and $M1$ strength functions and their possible contributions to the strength
function at very low energy are a challenging problem. 

In this Letter we  present shell-model calculations of the $M1$ strength
function. We study the isotopes $^{94}$Mo, $^{95}$Mo, $^{96}$Mo, in which the
low-energy enhancement has been observed, and the $N$ = 50 nuclide $^{90}$Zr.
A strong enhancement of the $M1$ transition strength is found below 2 MeV,
which accounts for the observed enhancement. The mechanism that generates the
strong low-energy $M1$ radiation will be explained.

The shell-model calculations were performed by means of the code RITSSCHIL
\cite{zwa85} using a model space composed of the
$\pi(0f_{5/2}, 1p_{3/2}, 1p_{1/2}, 0g_{9/2})$ proton orbits and the
$\nu(0g_{9/2}, 1d_{5/2}, 0g_{7/2})$ neutron orbits relative to a $^{68}$Ni
core. The configuration space was tested in detail in our earlier
shell-model studies of nuclei with $N = 46 - 54$ \cite{sch09,sch022,sch95,
sch982,sch06,win93,win94,rei95,ste00,zha04,jun98,jun99,ste01,rai02,ste02} and
was found appropriate for the description of level energies as well as
$M1$ and $E2$ transition strengths in nuclides around $A$ = 90. 
As a further test, we compared  the energies of yrast and yrare levels 
in $^{94,95,96}$Mo and $^{90}$Zr from the present calculation with the
experimental ones, which agree within 300 keV.

\begin{figure}
\vspace*{-6cm}
\hspace*{-1.5cm}
\epsfig{file=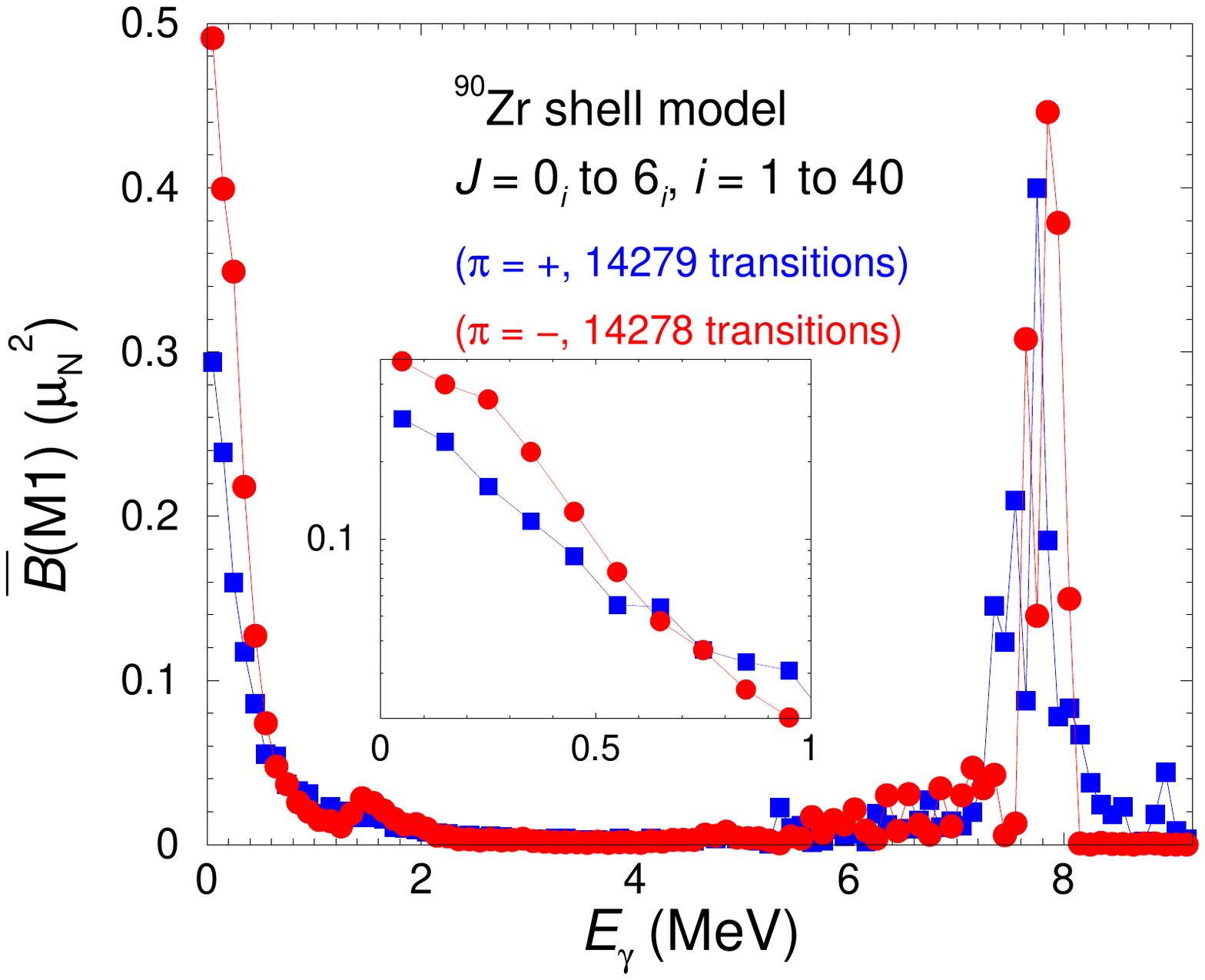,width=11cm}
\vspace*{-2.5cm}
\caption{\label{fig:90ZrEgM1}(Color online) Average $B(M1)$ values in 100 keV
bins of transition energy calculated for positive-parity (blue squares) and
negative-parity (red circles) states in $^{90}$Zr. The inset shows the
low-energy part in logarithmic scale.}
\end{figure}

\begin{figure}
\vspace*{-6cm}
\hspace*{-1.5cm}
\epsfig{file=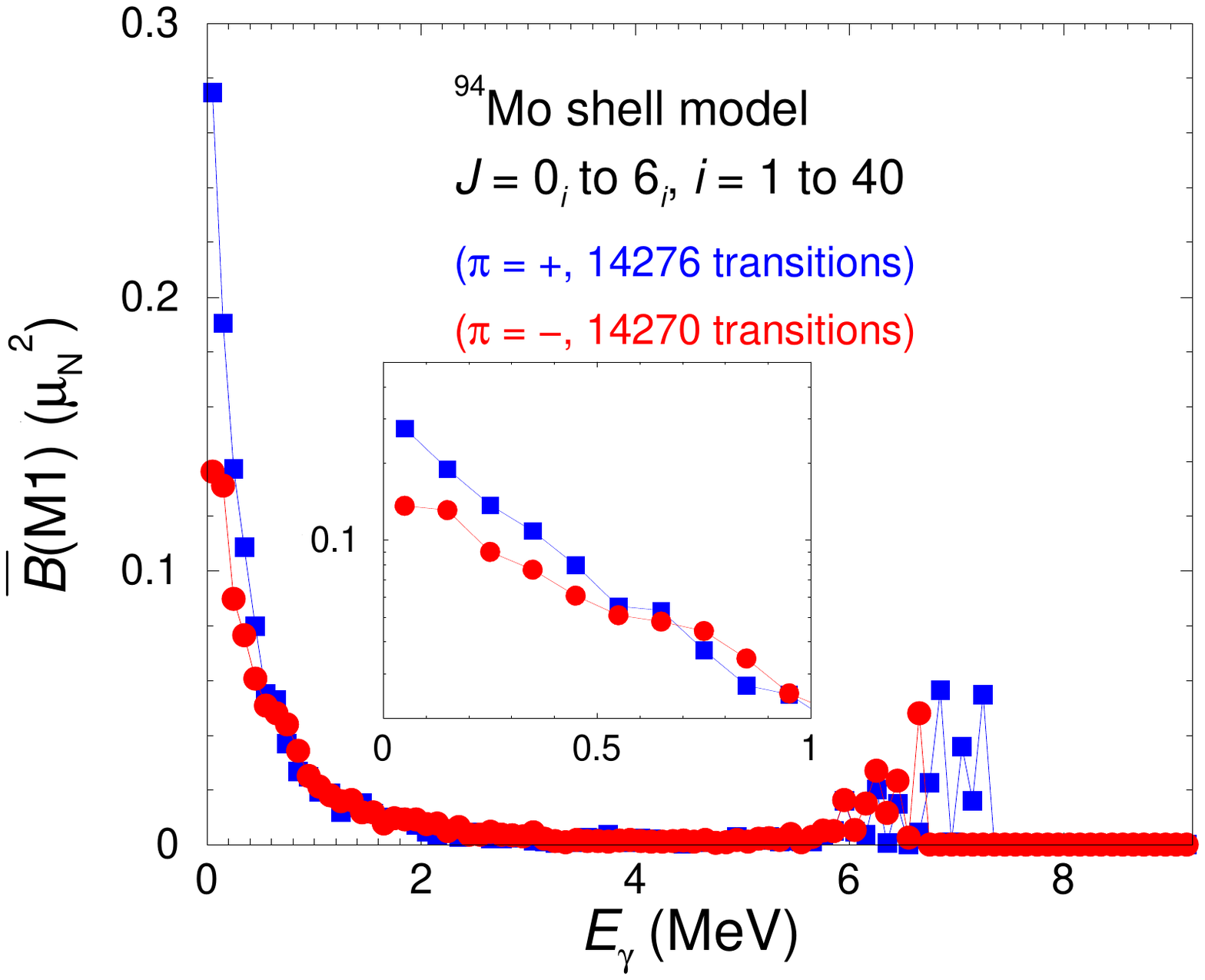,width=11cm}
\vspace*{-2.5cm}
\caption{\label{fig:94MoEgM1}(Color online) As Fig.~\ref{fig:90ZrEgM1}, but for
$^{94}$Mo.}
\end{figure}

The calculations included states with spins from $J$ = 0 to 6 for $^{90}$Zr,
$^{94}$Mo, $^{96}$Mo, and from $J$ = 1/2 to 13/2 for $^{95}$Mo. For each spin
the lowest 40 states were calculated. The reduced transition probabilities
$B(M1)$ were calculated for all transitions from initial to final states with
energies $E_f < E_i$ and spins $J_f = J_i, J_i \pm 1$. For the minimum and
maximum $J_i$, the cases $J_f = J_i - 1$ and $J_f = J_i + 1$, respectively,
were excluded. This resulted in more than 14000 $M1$ transitions for each
parity $\pi = +$ and $\pi = -$, which were sorted into 100 keV bins according
to their transition energy $E_\gamma = E_i - E_f$. The average $B(M1)$ value
for one energy bin was obtained as the sum of all $B(M1)$ values divided by the
number of transitions within this bin. The results for the nuclides $^{90}$Zr
and $^{94}$Mo are shown in Figs.~\ref{fig:90ZrEgM1} and \ref{fig:94MoEgM1},
respectively.

For all considered nuclides and each parity a pronounced low-energy enhancement
of the average $\overline{B}(M1)$ values is seen. The bump around 7 MeV in
$^{90}$Zr and $^{94}$Mo is caused by $1 \rightarrow 0$ and $0 \rightarrow 1$
transitions from states dominated by the spin-flip configuration
$\nu(0g_{9/2}^{-1} 0g_{7/2}^1)$. The cumulative strength calculated
for the $1^+ \rightarrow 0^+_1$ transitions in $^{90}$Zr is consistent with the
value deduced in a recent experiment as shown in Ref.~\cite{rus13}. In
$^{95}$Mo and $^{96}$Mo the bump around 7 MeV does not appear, because the 
excitation of a $1d_{5/2}$ neutron to the $0g_{7/2}$ orbit is
preferred to $\nu(0g_{9/2}^{-1} 0g_{7/2}^1)$.

The insets of Figs.~\ref{fig:90ZrEgM1} and \ref{fig:94MoEgM1} demonstrate that,
up to 2 MeV, the low-energy enhancement of $\overline{B}(M1,E_\gamma)$ is well
approximated by the exponential function
$\overline{B}(M1,E_\gamma) = B_0 \exp{(-E_\gamma/T_B)}$ with
$B_{0} = \overline{B}(M1,0)$ and $T_B$ being constants. For the respective
parities $(\pi = +, -)$ we find
for $^{90}$Zr: $B_0$ = (0.36, 0.58) $\mu^2_N$ and $T_B$ = (0.33, 0.29) MeV, 
for $^{94}$Mo: $B_0$ = (0.32, 0.16) $\mu^2_N$ and $T_B$ = (0.35, 0.51) MeV, 
for $^{95}$Mo: $B_0$ = (0.23, 0.12) $\mu^2_N$ and $T_B$ = (0.39, 0.58) MeV, and
for $^{96}$Mo: $B_0$ = (0.20, 0.13) $\mu^2_N$ and $T_B$ = (0.41, 0.50) MeV.

\begin{figure}
\vspace*{-7.3cm}
\hspace*{-0.5cm}
\epsfig{file=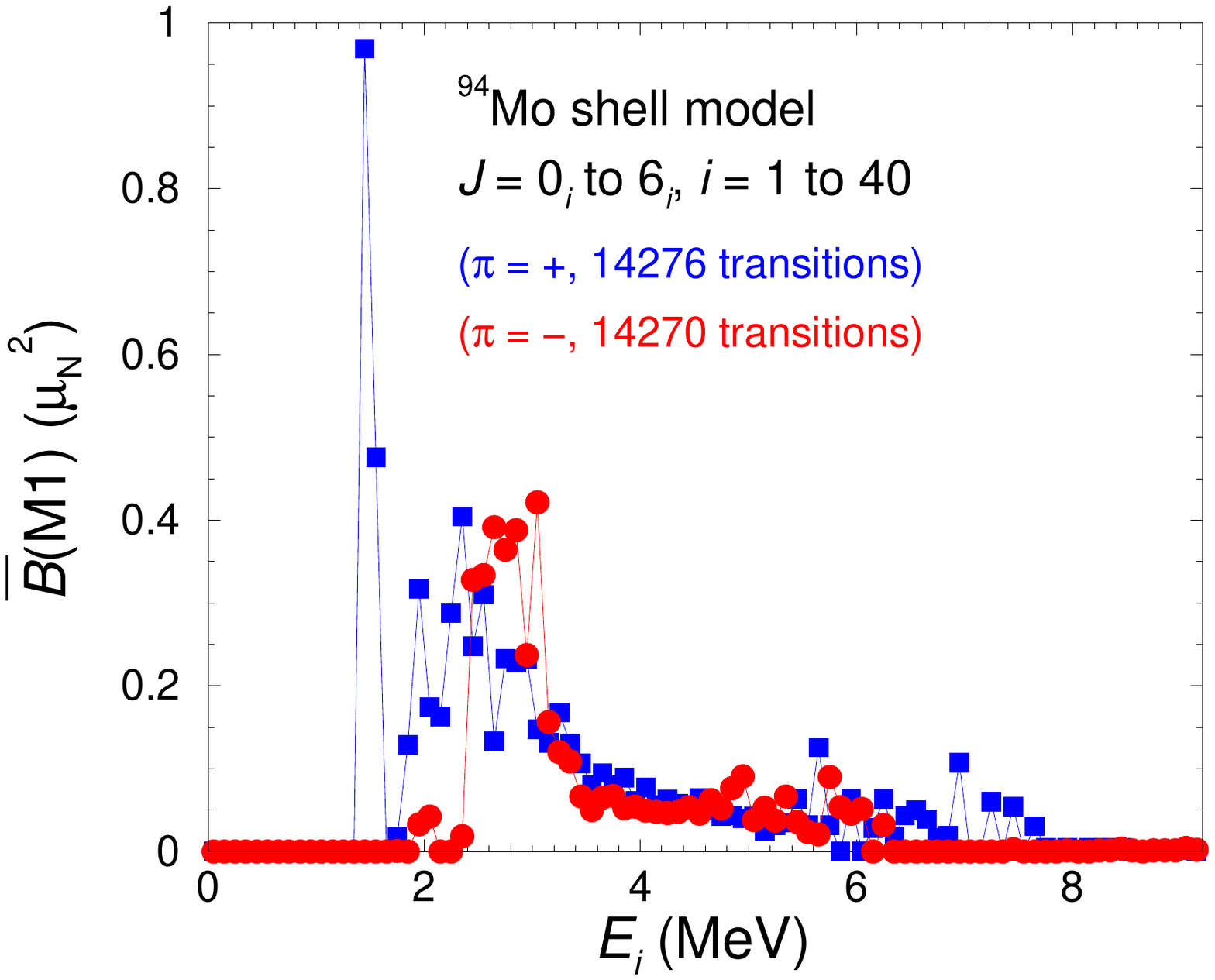,width=10.5cm}
\vspace*{-1cm}
\caption{\label{fig:94MoExM1}(Color online) Average $B(M1)$ values in 100 keV
bins of excitation energy calculated for positive-parity (blue squares) and
negative-parity (red circles) states in $^{94}$Mo.}
\end{figure}

To find out which states generate strong $M1$ transitions, the average
$\bar{B}(M1)$ values for $^{94}$Mo are plotted as a function of the energy of
the initial states in Fig.~\ref{fig:94MoExM1}. The large spike at 1.5 MeV
in the distribution of $\pi = +$ states arises from the
$2^+_2 \rightarrow 2^+_1$ and $4^+_2 \rightarrow 4^+_1$ transitions which 
link the main configurations $\nu(1d_{5/2}^2)$ in the $2^+_1$ and $4^+_1$
states with $\pi(0g_{9/2}^2) \nu(1d_{5/2}^2)$ in the $2^+_2$ and $4^+_2$
states. These findings are consistent with the experimental results given in
Refs.~\cite{pie99,fra03}, which list $E_\gamma$ = 1196 keV and
$B(M1) = 0.56(5) \mu_N^2$ for the corresponding $2^+_3 \rightarrow 2^+_1$
transition to be compared with the calculated values of 884 keV and 0.96 
$\mu_N^2$. The bump in the $\pi=+$ distribution between 2 and 3 MeV includes
among others  the $1^+_2$ state with the main configuration
$\pi(0g_{9/2}^2) \nu(1d_{5/2}^2)$. It deexcites with $B(M1) = 0.37 \mu^2_N$ to
the ground state, comparable with the experimental $1^+_2$ state described in
Refs.~\cite{pie99,fra03}. For $^{96}$Mo there are analogous similarities of the
calculations with the experimental results \cite{fra04,les07}. The broad
enhancements between 2 and 8 MeV (6 MeV) for the $\pi = +$ ($\pi = -$)
distributions contain contributions from many states, where all included
initial spins contribute approximately the same fraction. The
$\overline{B}(M1)$ distributions versus $E_i$ in $^{95}$Mo and $^{96}$Mo look
similar to the ones in $^{94}$Mo, but are shifted to somewhat lower excitation
energy. In $^{90}$Zr, the distributions start at about 3 MeV and continue to
10 MeV.

The low-energy enhancement of $M1$ strength is caused by transitions between
many close-lying states of all considered spins located well above the yrast
line in the transitional region to the quasi-continuum of nuclear states.  
Inspecting the wave functions, one finds large $B(M1)$ values for transitions
between states that contain a large component (up to about 50\%) of the same
configuration with broken pairs of both protons and neutrons in high-$j$
orbits. The largest $M1$ matrix elements connect configurations with the spins
of high-$j$ protons re-coupled with respect to those of high-$j$ neutrons to
the total spin $J_f = J_i, J_i \pm 1$. The main configurations are
$\pi(0g_{9/2}^2) \nu(1d_{5/2}^2)$, 
$\pi(0g_{9/2}^2) \nu(1d_{5/2}^1 0g_{7/2}^1)$, and 
$\pi(0g_{9/2}^2) \nu(1d_{5/2}^2 0g_{9/2}^{-1} 0g_{7/2}^1)$ for positive-parity
states in $^{94}$Mo. Negative-parity states contain a proton lifted from the
$1p_{1/2}$ to the $0g_{9/2}$ orbit in addition. In $^{90}$Zr, analogous
configurations are generated by exciting protons over the subshell gap at
$Z$ = 40 and neutrons over the shell gap at $N$ = 50, i.e. 
$\pi(1p_{1/2}^{-2} 0g_{9/2}^2) \nu(0g_{9/2}^{-1} 1d_{5/2}^1)$ and 
$\pi(1p_{1/2}^{-2} 0g_{9/2}^2) \nu(0g_{9/2}^{-1} 0g_{7/2}^1)$ for 
positive-parity states and only one $1p_{1/2}$ proton lifted for
negative-parity states. The orbits in these configurations have large $g$
factors with opposite signs for protons and neutrons. Combined with specific
relative phases of the proton and neutron partitions they cause large total
magnetic moments.

The $M1$ strength functions were deduced using the relation $f_{M1}(E_\gamma)
= 16\pi/9$ $(\hbar c)^{-3}$ $\overline{B}(M1,E_\gamma)$ $\rho(E_i)$.
They were calculated by multiplying the ${B(M1)}$ value in $\mu^2_N$ of each
transition with $11.5473 \times 10^{-9}$ times the level density at the energy
of the initial state $\rho(E_i)$ in MeV$^{-1}$ and deducing averages in energy
bins as done for the $\overline{B}(M1)$ values (see above). The level densities
$\rho(E_i,\pi)$ were determined by counting the calculated levels within energy
intervals of 1 MeV for the two parities separately. For the Mo isotopes, the
total level densities $\rho(E_i)$ are well reproduced by the
constant-temperature expression $\rho(E_i) = \rho_0 \exp{(E_i/T_\rho)}$ as long
as $E_i <$ 3 MeV. For higher energies the combinatorial level density deviates
from this expression and eventually decreases with excitation energy, which is
obviously due to missing levels at high energy in the present configuration
space. From a fit to the combinatorial values in the range $E_i < 2$ MeV we
found for $(\rho_0, T_\rho)$ in (MeV$^{-1}$, MeV) values of (1.37, 0.67),
(1.90, 0.54), and (1.25, 0.58) for $^{94}$Mo, $^{95}$Mo, and $^{95}$Mo,
respectively. The level density in the semi-magic $^{90}$Zr shows a more
complicated energy dependence. The total $M1$ strength functions for $^{94}$Mo,
$^{95}$Mo, and $^{96}$Mo are shown in Figs.~\ref{fig:94MoEgf1},
\ref{fig:95MoEgf1}, and \ref{fig:96MoEgf1}, respectively. As for the
$\overline{B}(M1)$, there is a pronounced enhancement below 2 MeV, which is
well described by the exponential function
$f_{M1}(E_\gamma) = f_0 \exp{(-E_\gamma/T_f)}$.
For $^{90}$Zr, $^{94}$Mo, $^{95}$Mo, and $^{96}$Mo, the parameters are
$f_0$ = (34, 37, 39, 55) $\times$ 10$^{-9}$ MeV$^{-3}$ and 
$T_f$ = (0.50, 0.50, 0.51, 0.48) MeV, respectively.

To compare the calculated strength functions with the ones deduced from the
($^3$He,$^3$He') experiments of Ref.~\cite{gut05}, the $E1$ contributions have
to be added. Because a calculation of the $E1$ strength within the present
model space is not possible, we adopted the GLO expression with parameters
adjusted to $(\gamma,n)$ data \cite{bei74} and the ($^3$He,$^3$He') data above
4 MeV, where our $M1$ contribution is negligible. In the comparison, we focus
on the low-energy region observed only via the ($^3$He,$^3$He') reaction,
whereas there exist also other experimental data for energies above about 4 MeV
\cite{rus09}. As seen in Figs.~\ref{fig:94MoEgf1}, \ref{fig:95MoEgf1}, and
\ref{fig:96MoEgf1}, the dipole strength functions found in the present
calculations resemble the ones deduced from $^3$He-induced reactions on
$^{93-98}$Mo \cite{gut05,lar10} and from a recent $^{94}$Mo$(d,p)^{95}$Mo
experiment \cite{wie12}, though the experimental data are available for
$E_\gamma >$ 1 MeV only. There is a certain freedom in determining the
parameters for the GLO, which results in some uncertainty of the magnitude of
the GLO in the enhancement region. As the GLO gives only a minor contribution
to the total strength below $E_\gamma$ = 2 MeV, an acceptable modification of
the parameters will not remove the exponential enhancement caused by the
$M1$ radiation. It will change the values around 2 MeV, leaving room for other
possible enhancement mechanisms. The comparison suggests that at least part of
the low-energy enhancement in the experimental dipole strength functions can be
explained by $M1$ transitions in the quasi-continuum of states. The analogous
low-energy enhancement predicted for $^{90}$Zr suggests an experimental study
of this nuclide.

Recent work \cite{lit13} suggested that thermal coupling of quasiparticles
to the continuum of unbound states may enhance the low-energy $E1$ strength.  
To account for the enhancement observed at 1 MeV, the temperature has to be
above 1.4 MeV. This temperature is higher than temperatures predicted by the
constant-temperature and Fermi-gas models (0.8 -- 0.9 MeV) \cite{egi09}, the
ones deduced from our shell-model level densities (0.6 MeV), and the
experimental values (0.8 -- 1.0 MeV) derived in Ref.~\cite{gut05}. In contrast
to our $M1$ strength function, which peaks at zero energy, the $E1$ strength
function has a maximum near 1 MeV and disappears at zero energy. The data in 
Figs.~\ref{fig:94MoEgf1}, \ref{fig:95MoEgf1}, and \ref{fig:96MoEgf1} are
compatible with a combination of both mechanisms, where the relative 
contribution to the low-energy enhancement cannot be assessed.

The re-coupling of spins leading to large $B(M1)$ values has been discussed in
connection with high-spin multiplets (see, e.g. 
Refs.~\cite{sch95,sch982,sch022,sch09}). An analogous mechanism generates
the ``shears bands'' manifesting ``magnetic rotation'' \cite{fraRMP}, which was
also observed in the mass-90 region \cite{sch99,sch022}. The ``mixed-symmetry''
configurations of the interacting boson model arise also from a reorientation
of the proton angular momentum with respect to the neutron one. All these
phenomena appear in nuclei near closed shells, if there are active high-$j$
proton and neutron orbits near the Fermi surface with magnetic moments adding
up coherently. Because these conditions are also prerequisites for the
low-energy enhancement, one may expect it to appear in the same nuclei as the
phenomena just mentioned. For example, the mixed-symmetry configurations
discussed for $^{94}$Mo \cite{pie99,fra03} and $^{96}$Mo \cite{fra04,les07}
correspond to the dominating configurations $\pi(0g_{9/2}^2) \nu(1d_{5/2}^x)$
($x$ = 2, 3, 4 for $^{94,95,96}$Mo, respectively) that were found causing large
$B(M1)$ strengths in the present calculations. The regions in the nuclear
chart, where magnetic rotation is expected, are delineated in Fig.~22 of
Ref.~\cite{fraRMP}. In fact, $^{90}$Zr and the Mo isotopes discussed in the
present work as well as the Fe, Ni, and Cd isotopes, for which the low-energy
enhancement was observed \cite{voi04,voi10,lar13}, belong to these regions. On
the other hand, $^{117}$Sn \cite{agv09}, $^{158}$Gd \cite{chy11}, and the Th,
Pa isotopes \cite{gut12}, for which no low-energy enhancement was observed, lie
outside these regions.
In Ref.~\cite{lar10} it was demonstrated that a low energy-energy enhancement
of the dipole strength function comparable with the present one for $f_{M1}$ 
(GLOup2 in Ref.~\cite{lar10}) increases the astrophysical $(n,\gamma)$ rate of
the r-process by more than a factor of 10. A comparable increase may be
expected for nuclei near the neutron drip line located in the mass regions
around $(Z,N)$ = (22,48), (26,52), (34,80), (64,118), where magnetic rotation
has been predicted and thus, the $M1$ strength should be enhanced at low
energy.  

Summarizing, the present shell-model calculations result in a large number of
low-energy $M1$ transitions between excited states. Their average strength
steeply increases toward zero transition energy. The strong radiation is
generated by a reorientation of the spins of high-$j$ proton and neutron
orbits. This $M1$ radiation accounts for the enhancement of the dipole strength
found in experiments. However, the uncertainties in calculating the low-energy 
$E1$ strength leave room for additional mechanisms.

We thank F. D\"onau, H. Grawe, E. Grosse, and E. Litvinova for stimulating
discussions. A. C. L. acknowledges support from the Research Council of Norway,
project grant no. 205528, and  S. F. acknowledges support by HZDR and by the
DoE Grant DE-FG02-95ER4093.

\end{document}